\documentstyle[12pt]{article}
\newcommand{\be}{\begin{equation}}
\newcommand{\ee}{\end{equation}}
\newcommand{\bear}{\begin{eqnarray}}
\newcommand{\eear}{\end{eqnarray}}
\newcommand{\ba}{\begin{array}}
\newcommand{\ea}{\end{array}}

\input math_macros
\newdimen\tdim
\tdim=\unitlength
\def\bar{\overline}

\begin{document}

\pagestyle{empty}
\begin{titlepage}
\def\thepage {}    

\title{  \bf  
Topology in the Bulk:\\
Gauge Field Solitons in Extra Dimensions \\ [2mm] } 
\author{
\bf  Christopher T. Hill$^{1}$ \\[2mm]
 and \\[2mm]
\bf Pierre Ramond$^2$ \\ [2mm]
 {\small {\it $^1$Fermi National Accelerator Laboratory}}\\
{\small {\it P.O. Box 500, Batavia, Illinois 60510, USA}}
\thanks{e-mail: hill@fnal.gov, ramond@phys.ufl.edu;}
\\ [3mm]
{\small {\it $^2$Institute for Fundamental Theory,}}\\ 
{\small {\it Physics Department, University of Florida}}\\
{\small {\it Gainesville, Florida, 32611}}\thanks{supported
in part by the Department
of Energy under grant DE-FG02-97ER41029}
}

\date{ }

\maketitle

\vspace*{-16.0cm}
\noindent

\begin{flushright}
FERMILAB-Pub-00/174-T \\ [1mm]
June 19, 2000 
\end{flushright}

\vspace*{14.1cm}
\baselineskip=18pt

\begin{abstract}

  {\normalsize
Certain 
static soliton configurations of gauge fields in $4+1$ dimensions
correspond to the
instanton in $4$ Euclidean dimensions ``turned on its side,''
becoming a monopole in $4+1$. 
The periodic
instanton solution can be used with
the method of images to construct solutions satisfying 
the brane
boundary conditions. The $\theta$-term
on the brane becomes a topological current source,
yielding an emission amplitude for monopoles into the bulk. 
Instantons have a novel reinterpretation
in terms of monopole exchange between branes.
} 
\end{abstract}

\vfill
\end{titlepage}

\baselineskip=18pt
\renewcommand{\arraystretch}{1.5}
\pagestyle{plain}
\setcounter{page}{1}


\section{Introduction}

The possibility that extra space-time dimensions
may be associated with lower energy scales,
$\sim 1$ TeV, rather than $M_{Planck}$ has
become an active area of discussion in the past few years.
Though many nagging theoretical questions 
abound in these schemes, e.g., ``how
do they match onto a string theory at higher
energies ?,'' they do potentially offer
some intriguing new possibilities for extensions
beyond the Standard Model, and would have
enormous significance for experiment at very high,
albeit accessible,
energies.

Models of extra-dimensions at the $\sim$ TeV scale
fall into two categories, those with gravity only in
the bulk [1,2], and those that include
gauge fields, as well as fermions, in the  bulk [3].
It has been argued
that the latter models can unify, in principle, at much lower
scales than 
the usual GUT scale, due to the dimensional coupling constant
and the accelerated power law running. 
In fact, this unification occurs generally for strong coupling
and we expect strong dynamics in these schemes at the
string scale [4].
 Many phenomenological
constraints
have been placed upon the scales of the new extra
dimension(s),
and a compactification
scale lower limit of order  $\sim 1$ TeV seems to be
indicated.

In the present paper we wish to discuss another 
feature of gauge fields in the bulk
which is potentially  
important to the phenomenology, and perhaps the viability
of these schemes.
Extra dimensions with bulk gauge fields
necessarily imply the existence of new
physical,  topologically stable
static soliton solutions.  The solitons we consider
presently are composed only of the
gauge degrees of freedom that are
extended to the bulk, and do not involve any
Higgs fields at high energies.  
We will confine our attention at present to
$4+1$ bulk space-time dimensions.

The salient new feature in extending beyond $3+1$
dimensions to, e.g., $4+1$, is that the homotopy
class $\Pi_3(SU(2))$, previously associated
with the instanton winding number, now becomes associated
with ``magnetic charge.''  While instantons
are quasi-localized ``events'' of finite action in
$3+1$ [5,6], the monopoles are static particles, ``world-lines'' 
in $4+1$, and have a mass determined by the compactification
scale.  (In $4+1$ dimensions the
gauge coupling constant has mass dimension
$-1/2$, and can be written as $g = g_0/\sqrt{M_c}$,
where $g_0$ is dimensionless;  $M_c$ is the compactification
scale.)

All nonabelian Yang-Mills gauge groups,
including those of the Standard Model, admit $SU(2)$
subgroups.
All $SU(2)$ gauge theories have instanton solutions,
when these theories are considered in Euclidean $d=4$.  These
are
``pseudoparticle solutions'' of the pure Euclidean gauge
equations
of motion, representing the homotopy group $\Pi_3(SU(2))$ [5], and
correspond physically to WKB tunneling solutions in the
quantum theory [6].  The tunneling is associated with the
violation
of a fermionic chiral 
charge, connected to the index of the instanton through
the axial anomaly.  
In the case of QCD this charge is the $U(1)_A$ axial
charge, and the instantons produce large tunneling amplitudes
near the QCD scale,
leading to an explicit
breaking of the chiral $U(1)_A$ symmetry and the removal of an
unwanted low mass Nambu--Goldstone boson from the spectrum [6].  
However, the overall $U(1)_A$ phase of the quark mass-matrix
is now elevated to a problem for QCD, the $\theta$ problem,
and to date the resolution of this problem is unknown.
In the case of
the electroweak $SU(2)_L$, the charge is $B+L$, and the 
instantons, which have very small amplitudes, 
lead to an unobservable low rate of proton decay.  Thus
instantons,
at least in QCD,
play an important role in the physics of the Standard Model.

If Standard Model gauge fields, either 
those of QCD or $SU(2)_L$, can
live in the bulk, we see immediately that there is a new
aspect of the usual instanton.  
The instanton, which was previously a pseudoparticle  
on a  Euclideanized $3+1$ brane, becomes a static, real
massive particle, 
in the bulk.  It becomes a magnetic monopole (though the
magnetic charge will not necessarily be electromagnetic;
in the QCD case it is chromomagnetic).
The two-form field strength is dual to a 3-form
in $4+1$.  However,
if we contract the three-form dual with the time-like vector
defining the solution's rest--frame, then there is a
self-duality,
the remnant of self-duality of the $d=4$ instanton. The
topological
charge is the space integral of a 4-form time component.
Self-duality implies that the mass of
the  monopole is therefore determined;
this object is a generalized
magnetic monopole, and shares many features in common
with the BPS monopole in $3+1$.

This new object
has a modulus, the parameter associated with
the size of the remnant instanton. If the theory
is scale invariant, then this quantity is undetermined,
and we have a one-parameter family of degenerate
solutions of monopoles varying sizes. 
In the scale-invariant limit the monopoles
have a common mass, $8\pi^2 M_c/g_0^2$
where $M_c$ is the compactification scale, and
$g_0$ a dimensionless gauge coupling.

An immediate complication arises when we wish
to consider monopoles in the bulk
bounded by branes. Gauge fields must
satisfy certain boundary conditions on the branes,
and we must obtain self-dual monopole 
solutions consistent with these.  Fortunately,
there exist exact self-dual 
solutions for multi-instantons on a line [7],
and these can be used with the method of images
to construct self-dual monopoles consistent with
the boundary conditions of the branes.
Again, self-duality is powerful, and the interaction
energy of the monopoles with the branes, or with each other,
is zero.

As an application, we consider the QCD $U(1)_A$
problem as viewed from a $4+1$ theory.
The $\theta$--term is not Lorentz invariant
in $4+1$ dimensions, and becomes a topological current
source from the brane into the
bulk, the analogue of a thermal ``cathode current''
in a vacuum tube.
Instantons still
exist in the theory,
appearing now as cylindrical solutions, 
analogous to vortices, independent
of $x^5$,
in the Euclidean 
$5$-dimensions that interconnect branes.  In QCD,
chiral charge can be removed from brane I by the 
instanton ``vortex'' and
transported to brane II.  
The instantons, however, can be reinterpreted as 
emission/absorption vertices 
for monopoles into the bulk (with
zero extant in $x^5$), and
the instanton vortex as the ``t-channel''
exchange of a monopole between branes.
The emission amplitude is now $\sim{\cal{O}}(1)$
rather than $\sim \exp(-8\pi^2/g_0^2)$.
The 't Hooft nonperturbative factor $\exp(-8\pi^2/g_0^2)$
is restored by
the  exchange of monopoles ``shining'' 
from one  brane to another.
This produces a Yukawa factor  $\exp(-M\delta)
= \exp(-8\pi^2/g_0^2)$, where $\delta$ is the
interbrane spacing. 

We construct an effective
monopole Lagrangian in the bulk.  
Monopoles couple to the branes via 
instantons, allowing the exchange of topological
charge between brane and monopoles in the bulk.  
We show how a nontrivial
monopole condensate forms between the
branes, which can support the topological
current. 
The $\theta$--potential is obtained
of the form $\propto \Lambda^4
\exp(-M\delta) \cos\theta$, which
represents the monopole condensate 
(``space charge'') response 
to the applied current.  

This picture bears some resemblance to the AdS-CFT 
correspondence
in string theory. 
The boundary theory of point-like,
or $0$-dimensional
instantons, is equivalent at low energies
to a bulk theory
of monopoles ($1$-dimensional world-lines).
It also suggests a dimensional
sequence of physical objects that
are correlated with anomalies, i.e.,
instanton in $d$ dimensions and
$\theta$-term becomes
a monopole in $d+1$ with $\theta$-current.

\section{$4+1$ Solitons based Upon Instantons}

The arena of this investigation is a  $d=4+1$ spacetime (the
bulk) in which two parallel branes, defining $d=3+1$ internal
worlds, are immersed. The ordinary 
spacetime coordinates are  labeled by $x^\mu$, $\mu=0,1,2,3$, 
and  the fifth dimension by $x^5$ (to avoid confusion with
$x^4=ict$).
The branes are respectively located at I: 
$x^5=\delta_I $ and II: $x^5= \delta_{II} $, with a constant 
interbrane separation $\delta =  \delta_{II} -  \delta_{I}$.

We first consider a pure $SU(2)$ gauge theory which lives in
the bulk, momentarily ignoring 
 fermions. It is defined by  the covariant
derivative $D_A = \partial_A + i A_A^a\tau^a/2$,  with  field
strengths $i\tau^a/2F^a_{AB} = [D_A, D_B]$, where capital
letters denote the bulk coordinates, $A,B=0,1,2,3,5$, and
$a,b=1,2,3$ are the $SU(2)$ labels. The canonical mass
dimension of the vector 
potential in $4+1$ dimensions is $3/2$, and the coupling
constant must have dimension
$-1/2$.  We introduce a scale $M_c$ and define the
scale invariant coupling constant $g\sqrt{M_c} \equiv g_0$.
$M_c$ becomes the compactification scale, and $g_0$ the
dimensionless coupling constant in
the effective $3+1$ theory at low energies.
Note that the dual
field strength in $4+1$ dimensions is the three-form:
\be
\widetilde{F}_{ABC}^a = \frac{1}{2} \epsilon_{ABCDE}F^{aDE}
\ee
Define $\chi_A$ to be a spacelike 5-vector normal to the
branes. Then a necessary
gauge-invariant prescription of the gauge field boundary
conditions on a brane 
is 

\be \chi_A F^{aAB} = 0\ ,\qquad {\rm at}~~ x^5=\delta_{I,II}\
.\ee
This removes  unwanted gauge invariant vector field strengths
in the $3+1$ theory.
The effective $3+1$ brane dual becomes $\widetilde{F}^{aAB} =
\chi_C \widetilde{F}^{aABC}$. The simplest gauge
choice realizing these boundary conditions is to impose
Neumann  conditions for 
$A^a_\mu$ with $\mu=0,1,2,3$, i.e.
$\partial A^a_\mu/ \partial x^5 = 0$, at $x_5 =
\delta_{I,II}$, and
Dirichlet conditions for the $3+1$ ``scalars'' $A^a_5$, i.e.
$A^a_5 = 0$ at $x_5 = \delta_{I,II}$.    The lowest energy
physical $A_i$ modes are zero-modes, independent of $x_5$, and
form the usual $3+1$ gauge field. The effective theory,
truncating of these
modes and integrating over $x_5$ yields the normal $3+1$
Lagrangian 
with  coupling constant $g_0$ and where $M_c\delta=1$. The
solutions with nontrivial $x_5$ 
dependence, including $A_5$, are Kaluzsa-Klein modes; hence
the boundary conditions neatly lift $A^a_5$ to a large mass,
which would otherwise remain as an unwanted
scalar excitation in the low energy $3+1$ theory. It follows
that at short distances in the bulk we have approximately a
continuum, scale invariant, $4+1$ dimensional gauge theory.

With nonabelian gauge fields in the bulk and two parallel
branes there arises the possibility of  topologically
nontrivial Wilson line
connections between the branes [8]. The branes are analogous to a
Josephson junction. The Wilson line phase for an $SU(2)$
theory can be topologically nontrivial throughout the
space-time
of the bulk:
\be
\Omega_{I,II} = P\exp\left(ig \int_I^{II} dx^5 A^a_5
\frac{\tau^a}{2} \right)
\ee
For example, for large $r$ in $3$ space dimensions on brane I
connecting to brane II, we can have:
\be
\Omega_{I,II} = \exp\left(i\pi  \frac{\tau^ix_i}{2r} \right)
\qquad r \rightarrow \infty\ .\ee
This can happen in the presence of a single soliton solution
as we now describe.

In $4+1$ dimensions,  consider the gauge-invariant 5-vector
\be
Q^A = \frac{g^2}{32\pi^2} \epsilon^{ABCDE} F_{BC}^a
{F}_{DE}^a\ .  \ee
It can be seen to satisfy 
\be
\partial_A Q^A=0\ ,
\ee
 leading  to the  conserved charge
\be
Q = 
\frac{g^2}{32\pi^2}\int d^4 x\;
\epsilon^{0\mu\nu\lambda\tau}
F_{\mu\nu}^a {F}_{\lambda\tau a}\ ,  
\ee
which must be associated with the soliton solution we are
seeking.

We first consider the limit of very 
large separation to the branes, $\delta_I=-\infty$ and
$\delta_{II} \rightarrow \infty$, where the boundary
conditions become irrelevant. In that case, we can
write the isolated soliton solution immediately, using 't
Hooft's original
form for the instanton, by interpreting the Euclidean $x_4$ in
his formula as $x_5$, thus producing a static finite energy
monopole solution of the form:
\begin{eqnarray}
\label{mon1}
A_i^a & = & \frac{2}{g}\left[ 
\frac{ \epsilon_{a ij}   
[x-r]^j  + \delta_{ai}[x^5-r^5] } {[x-r]^2 + \rho^2} 
\right];
\qquad (a, i, j) = 1,2,3\ .
\nonumber \\
A_5^a & = & -\frac{2}{g}  
\frac{[x-r]^a}{[x-r]^2 + \rho^2}\ ; 
\nonumber \\
A_0^a & = & 0\ , 
\end{eqnarray}
where 
$[x-r]^2$ (square brackets) should be
understood as
\be
[x-r]^2 = (\eta^A\cdot(x-r)_A)^2 -(x-r)^A(x-r)_A\ ,
\ee
and $\eta^A$ is  the monopole $5$-velocity. The
solution is
static, centered at $r$ (the 
square bracket $[x - r]$ are synchronous, i.e.,
$x_0=r_0$), and 
we have fixed a particular gauge orientation
of this solution, where the hedgehog component
is chosen to be $A_5^a$. 

This solution is ``self-dual'' in the sense that:
\be
\label{eq2.7}
\eta^A\widetilde{F}_{ABC}^a = F_{BC}^a\ ,
\ee
so that the integer--valued topological index is the spatial
integral of the charge
density of the current $Q_A$ in the monopole rest-frame.

This solution can be written in the more compact form,
following [9]:
\be
\label{ext}
A_A^a\frac{\tau^a}{2}
= \frac{2}{g}\bar{\sigma}_{AB} \Pi^{-1}\partial_B \Pi
\qquad\qquad
\Pi = \left([x-r]^2 + {\rho^2}\right)
\ee
where the $\bar{\sigma}_{AB}$ are antisymmetric and are
related to the
`t Hooft matrices as:
\begin{eqnarray}
\bar{\sigma}_{ij} & = &  
\epsilon_{aij}\frac{\tau^a}{2i} \qquad \qquad \qquad  a,i,j =
1,2,3\ ,
\nonumber \\
\bar{\sigma}_{5i}&=& -\frac{\tau^i}{2i};
\qquad \qquad  \bar{\sigma}_{0A} =0\ .
\end{eqnarray}
In the form for $\Pi$ of eq.(\ref{ext})
the monopole charge,
which is the $4$-volume integral over $Q_0$, the ``Pontryagin
index," can be
written as the surface integral at infinity over a
gauge-dependent current.
One can write a gauge equivalent solution\footnote{This
involves a conformal inversion, as well, and is not 
gauge equivalent if conformal breaking effects,
e.g., quantum loops, are considered.}
multiplying $\Pi$ by $[x-r]^{-2}$, which
does not change the action or the global Pontryagin
index, hence:
\be
\label{eq11}
A_A^a\frac{\tau^a}{2}
= \frac{2}{g}\bar{\sigma}_{AB} \Pi^{-1}\partial_B \Pi
\qquad\qquad
\Pi = \left( 1+ \frac{\rho^2}{[x-r]^2} \right)
\ee
This latter solution should be viewed as  ``punctured,''
in that the singularities are not integrated; the Pontryagin
index is given by
the small surface integral surrounding the singularity, and
the
surface at infinity gives no contribution to the charge.
This form of the solution is, in a sense, a localizeable
monopole, localized at $r$ of size $\rho$. This latter form is
most useful for generalization to  solutions with the proper
brane boundary conditions. 
 
The energy for this static solution is given by the
usual expression for the action of a $3+1$ instanton:
\be
S = -\frac{1}{4}
\int d{x}^4 \; F_{AB}^a F^a_{AB} = -\frac{8\pi^2 M_c}{g_0^2}
\ee
where  $F_{AB}^a $ are the field strengths. While this
expression was the Euclidean action for the instanton, it now
plays the role of the mass of our static solution.
Hence, this describes a static, topologically
stable, magnetic monopole solution with a mass of $8\pi^2
M_c/g_0^2$. 
and thus represents the nontrivial $\Pi_3(SU(2))$. This is the
relevant homotopy for
a monopole in $4+1$ dimensions, the analogue of
$\Pi_2(G/U(1))$ in $3+1$ dimensions.

The monopole solutions of eq.(\ref{mon1}, 
\ref{ext}, \ref{eq11}) 
are only consistent with the boundary conditions in the large
$\delta$ limit.
Fortunately, one can readily generalize these to
multi--monopole solutions along an infinite $x^5$ line [7,8].
This allows us to use the method of images to implement the
boundary conditions.

Consider, first for simplicity, brane I to be at the origin,
$x^5=0$, hence
$\delta_I=0$ and brane II at infinity, $\delta_{II} =\delta
\rightarrow \infty$.
We locate a monopole at $r = (r^0,0,0,0, r^5)$. We must have
that $A_5(x^5) = -A_5(-x^5)$ and $A_i(x^5) = A_i(-x^5)$ to
satisfy the gauge field boundary conditions described above.
Thus, we need to place the appropriate {\em image monopole }
at $r' = (r^0,0,0,0, -r^5)$, with identical gauge orientation.
These conditions are satisfied if we require
$\partial_5\Pi =0$ when $x^5=0$. We thus choose:
\be
A_A^a\frac{\tau^a}{2}
= \frac{2}{g} \bar{\sigma}_{AB}\Pi^{-1}\partial_B \Pi
\qquad\qquad
\Pi = \left( 1 + \frac{\rho^2}{ [(x-r)]^2 }  + \frac{\rho^2}{
[(x-r')]^2 }  
\right)
\ee
This self-dual solution represents two identical monopoles,
valid everywhere,
through the physical region $x^5>0$.

We now bring in the second brane from infinity, so that 
$\delta_{II} =\delta<\infty $. With the source monopole at 
$r = (r^0,0,0,0, r^5)$, we now require an infinite set of
image monopoles at $r_n' = (r^0,0,0,0,-r^5 + 2n\delta)$, 
and $r_n = (r^0,0,0,0, r^5 + 2n\delta)$, all with identical
gauge orientations (see Fig.(1)). Hence:
\be
A_A^a\frac{\tau^a}{2}
= \frac{2}{g} \bar{\sigma}_{AB} \Pi^{-1}\partial_B \Pi\ ;
\qquad\qquad
\Pi =  
\left( 1 + \sum_{n=-\infty}^\infty
\frac{\rho^2}{ [x-r_n]^2 }  + 
\sum_{n=-\infty}^\infty
\frac{\rho^2}{ [x-r_n']^2 }  
\right)
\ee
%
%
\begin{figure}[t]
\vspace{7cm}
\includegraphics{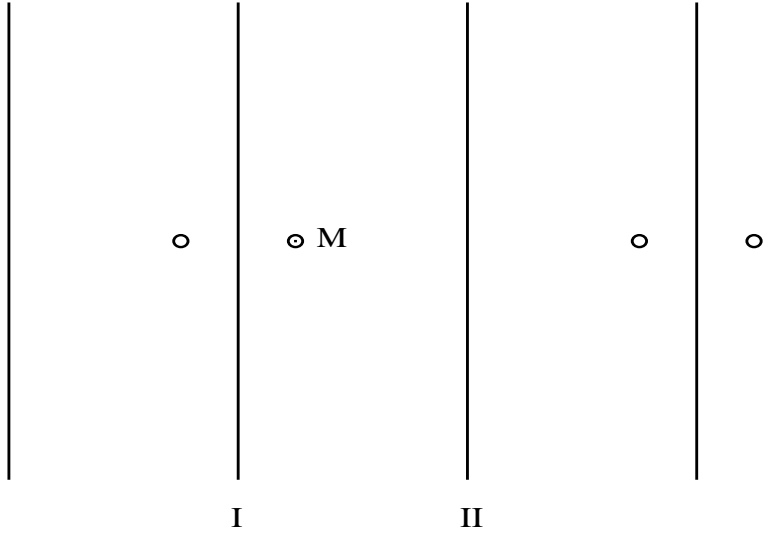}
\vspace{1cm}
\caption[]{
The monopole and its nearest neighbor image domains.
The solution is self-dual, following from the Witten solution
for
an infinite number of instantons on a line. Self--duality
implies
there is no dependence upon $r^5$ of the monopole energy.}
\end{figure}

\noindent The solution is self-dual in the sense of
eq.(\ref{eq2.7}).
This insures that the energy is independent of $r_5$,
i.e., 
at this level there is no nontrivial interaction potential
with the brane (we will see that this is modified by
instantons
below).
The boundary conditions by themselves
imply that a monopole 
bounces off the brane with unit S-matrix. 
A brane--monopole collision can be viewed
as the exchange of the monopole
with its image. 
For such a bounce off the brane the net
change in $\int_I d^4x F\widetilde{F} =0 $ since
the orientation of the surface integral over
the right monopole upon contact with the brane
is opposite to that on the left.

Since the  $4+1$ quantum theory is nonrenormalizable,
it must be viewed as a cut-off field theory, implying strong
cut-off dependence of the 
 physical quantities (one of the nagging questions posed
in the introduction).
Normally, one views the cut-off in extra-dimensional models
as a finite limit on the KK-mode sum.  How does this
regularization affect the monopole solution we have described?
The solution involves Fourier components of arbitrarily
large momentum, and one must in
principle, discretize these with an upper
bound $ < M_s$, where $M_s$ is the ``string scale,''
and typically $M_c << M_s$. 

The presence of $M_s$ obstructs the inversion
transformation used to arrive at the periodic
solution. The solution cannot be valid on distance
scales $|x-r|< 1/M_s$.
The large-distance topological behavior remains intact,
so solutions such as eq.(\ref{eq11}), which
enforce the
topological charge as the surface integral at
infinity, remain valid.

We would expect that
the string--cut-off places a lower
bound on, and discretizes, the physical
sizes of our solutions with $\rho > 1/M_s$.
Moreover, the scale breaking
implies that the monopole will have
a prefered mass that is dependent upon the 
size modulus.  If the high energy theory
has a strong coupling behavior, we
would expect stable monopoles with 
$\rho \sim 1/M_s$, and $M \sim M_c$,
since $8\pi^2/g_0^2(M_s) \leq 1$.
We also expect stable monopoles with
$\rho \sim 1/\Lambda_{QCD}$, and $M \sim M_c$,
since $8\pi^2/g_0^2(\Lambda_{QCD}) \leq 1$.

 Let us now consider what happens
when we incorporate fermions as modes 
confined to exist only on brane I.
On the brane the axial current satisfies the anomaly
relation:
\be
\partial_\mu j^{\mu A} = \frac{Ng^2}{32\pi^2} F^a_{\mu\nu}
\widetilde{F}^{a\mu\nu}
\ee
For a Euclidean 4-volume we have Stokes theorem:
\be
\int_V d^4 x \; \partial_\mu j^{\mu A} = \int_{\partial V} d
{\cal{A}}_3^\mu j^A_{\mu }
\ee
where ${\cal{A}}_3$ is the 3-surface bounding the
Euclidean 4-volume on the brane.
Now, the anomaly is the fifth component of the
Q-current
flowing off the brane:
\be
\frac{Ng^2}{32\pi^2} F^a_{\mu\nu} \widetilde{F}^{a\mu\nu} 
= N\chi^A Q_A\ , 
\ee
hence:
\be
\int_{\partial V} d {\cal{A}}_3^\mu j^A_{\mu } =N \int_V d^4 x
\; 
 \chi^A Q_A 
= N\int_V d^5 x \; 
 \partial^A Q_A - 
 N\int^{'}_{\partial V} d {\cal{A}}_4^A \; Q_A
\ee
the latter integral on the {\em rhs} is over the
4-area bounding a 5-volume, excluding the patch
on the brane.  The minus sign is due to the sense
of the area pointing out of the 5-volume. 
Using the conservation of $Q$ we thus have:
\be
\label{eq221}
\int_{\partial V} d {\cal{A}}_3^\mu j^5_{\mu } =  
 - N\int^{'}_{\partial V} d {\cal{A}}_4^A \; Q_A
\ee 
This  says that axial current can flow into a region
of the brane and disappear. It becomes a charge
associated with the monopole off the brane.  This is
mediated by the presence of a nonzero expectation value of 
$F\tilde{F}$ on the brane.  This leads to
the
interpretation of instantons as emission 
and absorption vertices for monopoles. The effective
Lagrangian for the monopole, treated as a quantum particle,
together with the brane-instanton interactions will give
a modified picture of the $U(1)$ and $\theta$ problems.

\section{Instantons, $\theta$-term, and $U(1)$ Problem}

In the previous section, we have demonstrated the existence of 
monopole-like solutions of the Yang-Mills field equations in
the bulk. Our picture also gives a reinterpretation of the
instantons of the $3+1$ theory. 

The traditional instanton
remains as a Euclidean solution with the supplemental
condition that $A_5^a=0$, and the solution is independent of
$x^5$.  Hence, it immediately satisfies the brane boundary
conditions. 
If we pass to Euclidean time, $x^0\rightarrow i\tilde{x}^0$,
the solution can be viewed 
as a one-dimensional Euclidean ``vortex,'' which emanates from
brane I and terminates on brane II. 
The instanton removes one unit of the associated charge
from brane I and deposits it on brane II
as shown in Fig.(2).
Deformed solutions, as in Fig.(3) must exist
also, and must be
continuously 
connected energetically to the straight line vortex limit.
The mixing of $x^5$ with $\tilde{x}^0$ under deformation
implies that
the instanton can be viewed as a vertex for emission
of the monopole from the brane as in Fig.(3). The
monopole
in the Euclideanized time situation is identical to that in
the Minkowski situation, and the monopole rotated in 
Euclidean space-time becomes the instanton.

\begin{figure}[t]
\vspace{7cm}
\includegraphics{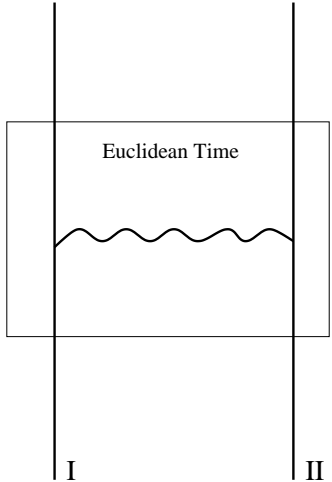}
\vspace{1cm}
\caption[]{
The traditional instanton is a cylindrical
solution, analogous to a vortex,
independent of $x^5$, in $4+1$ Euclideanized
 space-time.}
\end{figure}

If we cut through an instanton vortex exchanged between the
branes as in Fig.(4) we see that we have cut through the
exchange propagator of a monopole. Hence, the traditional
instanton in $3+1$ now appears as a ``t-channel'' exchange of
the monopole between the two branes
in $4+1$.  This picture gives us a dual interpretation
of instanton physics:
Whereas the amplitude for the instanton in Euclideanized $3+1$
takes
the form $\sim \exp(-8\pi^2/g_0^2)$, the t-channel monopole
exchange in $4+1$ involves a Yukawa factor $\sim
\exp(-M\delta) = \exp(-8\pi^2/g_0^2)$.
This reinterpretation can provide, as
we see below, an equivalent resolution of
$U(1)_A$ problem. 
An overall CP-violating $\theta$ angle
still remains in the present case. 

The emission vertex must be consistent with the
description of a leakage of current off brane I into
the bulk. The $\theta$-term in
$4+1$ is no longer Lorentz invariant,
and can be viewed as the current source
from the brane into the bulk.
In the presence of this source, the monopole
produces a coherent classical field between the branes.
This field, produced on one,
brane is exponentially damped as
$\sim \exp(-M\delta)$ as it ``shines'' on
the other brane, which reproduces the traditional instanton
tunneling probability,
$\sim \exp(-8\pi^2/g_0^2)$.
While the monopole field allows us to remove, by phase
redefinition,
any $\theta_i$ term on brane i, there remains
a net $\theta_{II} -\theta_I \equiv \theta$ term overall
in the physics, and the overall $\theta$ angle
remains. The monopole t-channel exchange 
reproduces the effect of the traditional instanton.
The ultimate removal of the residual $\theta$ term 
would require, e.g., an axion 
(which may arise on brane II in isolation from fermionic
structure on brane I).

\begin{figure}[t]
\vspace{7cm}
\includegraphics{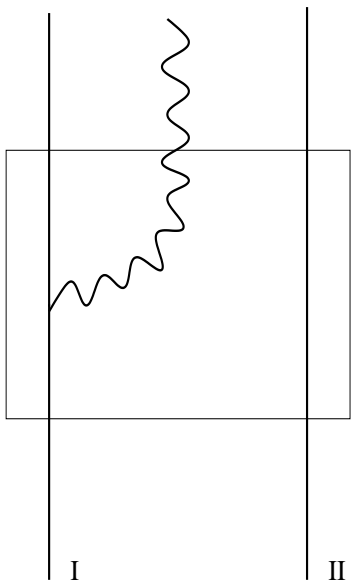}
\vspace{1cm}
\caption[]{
Instanton as emission vertex for a monopole off the brane into
the bulk.}
\end{figure}

\begin{figure}[t]
\vspace{7cm}
\includegraphics{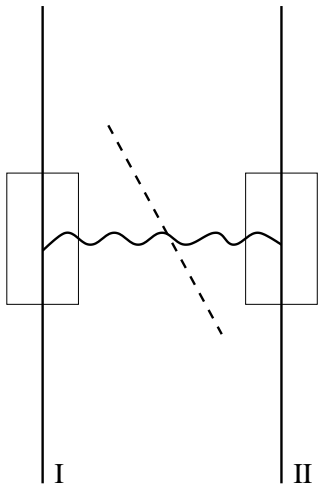}
\vspace{1cm}
\caption[]{
Cutting through instanton reveals 
monopole exchange.  The emission vertices can be
shrunk to localized (localized in $x^5$, of extent
$\epsilon$) Euclidean patches on the brane,
the $\exp(-8\pi^2 M_c\epsilon/g_0^2)$ 
tunneling amplitude for a patch approaching unity. The full
$\exp(-8\pi^2/g_0^2)$ tunneling amplitude factor is then restored
by the Yukawa 
exchange amplitude of the monopoles, $\propto \exp(-M\delta)$.
}
\end{figure}

A quantitative description of this picture
involves setting up 
an effective Lagrangian  describing the brane-monopole
dynamics.  
To that effect, we  represent the monopole 
by a complex scalar
field $F$ in the bulk. 
Although it should be parameterized by various moduli, 
including the size modulus $\rho$, we suppress these
collective parameters at present (we are considering
monopoles of size $\sim 1/\Lambda_{QCD}$.
We start with an  effective $4+1$ free Lagrangian density
in the bulk:
\be
\label{Lbulk}
{\cal L}_F =  
\partial_AF^\dagger \partial^A F  - M^2 F^\dagger F  \ .
\ee
The $x^5$--periodic gauge--field monopole solution 
implies supplemental periodic boundary conditions
on $F$:   $F$ must
be symmetric under 
$x^5\rightarrow 2\delta_m - x^5$, for a brane located at
$x^5=\delta_m$. Note that $F$ does not satisfy Neuman conditions,
since the field essentially parametrizes the centers of gauge
solitons. The current:
\be
J^A = \frac{i}{2} F^\dagger 
\stackrel{\leftrightarrow}{\partial}{}^A F \sim Q^A
\ee
must be identified with the topological current, $Q^A$.
This current is conserved in the bulk, but will have
sources at the branes since
our effective Lagrangian must represent eq.(\ref{eq221}),
i.e.,  the change in the $Q_5$ current component
upon traversing the brane must be given by the current source
term:
\be
\left. \frac{i}{2} F^\dagger 
\stackrel{\leftrightarrow}{\partial}{}^5 F \right|_{brane+}
-
\left. \frac{i}{2} F^\dagger 
\stackrel{\leftrightarrow}{\partial}{}^5 F \right|_{brane-}
=\;
<{\frac{Ng^2}{32\pi^2} F^a_{\mu\nu} \widetilde{F}^{a\mu\nu}} >
\ee
Note that, in QCD:
\be
<{\frac{g^2}{32\pi^2} F^a_{\mu\nu} \widetilde{F}^{a\mu\nu}} >
\; =
\frac{\delta }{\delta \theta } V(\theta)
\approx -\tilde{\kappa } e^{-8\pi^2/g_0^2} \Lambda_{QCD}^4\sin\theta
\ee
where $V(\theta)$ is the 
the $\theta$--potential (this is
essentially
$i\hbar \ln $ (Feynman path integral) in
the presence of instantons).  
Here $\tilde{\kappa}$ is the effective coupling constant
involving the full instanton configuration sum,
the determinant over small oscillations, etc.
In order to describe these effects, we must supplement the
free effective action for $F$ with 
a source term that describes
the monopole emission from the brane into
the bulk.

In the $4+1$ theory there is no Lorentz invariant
$\theta$-term. The $\theta$-term physically
depends upon the definition of the branes,
e.g., their fermionic
content, etc. 
For example, if fermions are localized on brane I, we must
allow for a localized
$\theta$ term on the brane, which appears as a ``topological
current'' source:
\be
{\cal L}_{\theta} = \frac{\theta_1Ng_0^2}{32\pi^2} 
\;\int d^5x \; \delta(x^5) \chi^{A}
\epsilon_{ABCDE} F^{aBC}F^{aDE}
\ee
where $\chi^{A}$ is the spacelike $5$-vector normal to the
brane pointing into
the bulk. Alternatively, we can view the  brane as a classical
object with an orientation  defined by $\chi_A$, and this is
simply an allowed  effective term
\footnote{
We'll conventionally put the full Standard Model fermion
structure on brane I. Alternatively,
with fermions in the bulk, which we do not
consider presently, we must
allow for a more general bulk-filling
$\theta$ term:
\be
{\cal L}_{Bulk}= \theta \;\int d^5x \; M_c \chi^{A}
\epsilon_{ABCDE} F^{aBC}F^{aDE}
\ee
This latter term will imitate a $\theta$-term in the effective
low energy $3+1$ theory, when appropriate boundary conditions
are implemented.
We will restrict our attention to the case of ${\cal
L}_\theta$. }.

In order to reproduce the essential current algebra embodied
in eq.(3.24), we need to include the amplitude for emission of
monopoles from the branes. This term should be proportional to
$e^{i\theta_i}$. Shifting in QCD 
$\theta_i\rightarrow \theta_i + \lambda$ 
generates the divergence of the axial current; 
the corresponding shift in 
$F\rightarrow \exp(-i\lambda) F$ generates the 
divergence of the $Q^A$ Noether current.
This leads us to postulate the  amplitude for emission of
monopoles from the brane:
\be 
{\cal L}_{\rm emission} =~\sum_{m=I,II}{\delta(x^5-\delta_m)}
\Lambda_{QCD}^2
M^{1/2} \tilde{g}_m
e^{i\theta_m} F + h.c.
\ee
where $\tilde{g}$ is a dimensionless parameter, and $\theta_m$
the $m$th brane's $\theta$ angle. The dimensionality of the
coefficient is chosen to match the dimensionality
of the $F$ field ($3/2$) and to satisfy consistency with the
instanton result. The $\sqrt{M}$ factor is a normalization
factor in $F$ consistent with the
proper normalization of a single quantum of $F$ carrying a
unit topological charge.

How would we compute $\tilde{g}_m$ from first principles? 
We give only a heuristic argument presently.
The emission vertex is an instanton patch on
the brane of finite length
$\epsilon$ in space, $x^5$,
i.e., the instanton, which is emitted from brane I, 
terminates on the monopole at $x^5=\epsilon$. The
nonperturbative
factor in the action for this is $\sim \exp(\int_0^\epsilon
dx^5\; 8\pi^2/g^2)
\sim \exp( 8\epsilon M_c \pi^2/g_0^2)$. This approaches
unity for $\epsilon \rightarrow 0$. Since we
are exchanging a large monopole, the relevant gauge coupling
constant feels large transverse distances in
the Euclidean $3+1$ dimensions, and is presumably
not suppressed by asymptotic freedom as
$\epsilon \rightarrow 0$. 
The usual instanton amplitude
prefactor, however, consisting of the bosonic determinant
for small oscillations around the instanton solution, and
a 't Hooft determinant over spin$-1/2$ fields,
would still be present.   Hence, $\tilde{g}_m$ is computed
as if we were calculating the action for an
instanton, but we discard the nonperturbative factor. 
The factor is local to the $m$th brane,
as it must be because the fermionic
content of a given brane is arbitrary.
The nonperturbative 
factor, as we stated above, returns through the $t$-channel
exchange of the monopole.  For
the emission of small monopoles, or on--shell
monopoles, we require a smaller transverse instanton,
and the relevant coupling constant $g_0$ becomes
$g_0(\rho)$. 

Neglecting fermions temporarily, we consider the effective
action
\be
{\cal L}_T={\cal L}_F+{\cal L}_{\rm emission}\ ,\ee
from which we derive the equation of motion for the monopole
field:
\be
0 = \partial^A\partial_A F + M^2 F - \sum_{m=I,II} \tilde{g}_m
e^{-i\theta_m} 
\Lambda_{QCD}^2 M^{1/2} \delta(x^5-\delta_m) 
\ee
We seek a solution is of the form:
\begin{eqnarray}
{F} & = & \alpha \exp(-M x^5) + \beta \exp( -M(\delta -
x^5)) \qquad 0 <
x^5 \leq \delta;
\nonumber \\
{F} & = & \alpha \exp(M x^5) + \beta \exp( -M(\delta +
x^5))  \qquad -\delta < x^5
\leq 0;
\nonumber \\
{F} & = & \alpha \exp(-M(2\delta - x^5)) + \beta \exp(
M(\delta - x^5))  \qquad
\delta \leq x^5
< 2\delta\ ,
\end{eqnarray}
where brane I has been set at $x^5=0$ and brane II at
$x^5=\delta$.\footnote{ We remark that we could allow an
overall phase factor, $\exp(i\omega/f_\omega)F$,
where $\omega$ depends upon $x^\mu$ and acts like a
(pseudo) Nambu-Goldstone boson; however, the 
boundary conditions, eq.(\ref{bound}), force $\omega =0$,
and small oscillations in $\omega$ are not relevant
to the physics in this case (see discussion below eq.(3.42)).}

This solution implements the even periodicity of
the underlying monopole gauge field solution,
which implies reflection symmetries about the branes.
The solution is infinitely periodic
as dictated by the infinite system of images,
however, we require it only
in these three domains to obtain $\alpha$ and $\beta$.
$\alpha$ and $\beta$ are determined by integrating
through the source delta-functions:
\begin{eqnarray}
\label{bound}
\left.\frac{\partial {F}}{\partial x^5}\right|_{x^5
=0^+}
-
\left.\frac{\partial {F}}{\partial x^5}\right|_{x^5
=0^-}
& = & -\tilde{g}_1\Lambda^2_{QCD} M^{1/2} e^{-i\theta_1};
\nonumber \\
\left.\frac{\partial {F}}{\partial x^5}\right|_{x^5
=\delta^+}
-
\left.\frac{\partial {F}}{\partial x^5}\right|_{x^5
=\delta^-}
& = & -\tilde{g}_2\Lambda^2_{QCD} M^{1/2} e^{-i\theta_2};
\end{eqnarray}
Hence, we have:
\begin{eqnarray}
\alpha & = &
\frac{\Lambda^2_{QCD}e^{-i\theta_1}}{4M^{1/2}\sinh(M\delta) }
\left(\tilde{g}_1e^{M\delta} + \tilde{g}_2 e^{i\theta}
\right);
\nonumber \\
\beta & = &
\frac{\Lambda^2_{QCD}e^{-i\theta_1}}{4M^{1/2}\sinh(M\delta)
}
\left(\tilde{g}_1 + \tilde{g}_2 e^{i\theta+M\delta}  \right)
;
\end{eqnarray}
where:
\be
\theta \equiv \theta_1 - \theta_2 .
\ee
The full solution is therefore:
\begin{eqnarray}
{F} & = & \frac{\Lambda^2_{QCD}e^{-i\theta_1}}{4M^{1/2}\sinh(M\delta)
}
\left[\left(\tilde{g}_1e^{M\delta} + \tilde{g}_2 e^{i\theta} 
\right)e^{-Mx^5}
+
\left(\tilde{g}_1e^{-M\delta}  + \tilde{g}_2 e^{i\theta}
\right)e^{M x^5}\right]
\end{eqnarray}
Note that the overall factor 
of $e^{-i\theta_1} $ is irrelevant,
and can be absorbed into a redefinition of ${F}$, and
only the net $\theta = \theta_1 - \theta_2$ remains in the physics.
This solution yields the nontrivial profile of the monopole
effective field $F$ between two branes with arbitrary 
$\theta$ terms.  It may be viewed as the manifestation of the
vacuum
condensate of monopoles between the branes.

We can readily verify the boundary condition on the current,
e.g., at brane I, $x^5=0$:
\be
\label{338}
\left. \frac{i}{2} F^\dagger 
\stackrel{\leftrightarrow}{\partial}{}^5  F \right|_{0+}
-
\left. \frac{i}{2} F^\dagger 
\stackrel{\leftrightarrow}{\partial}{}^5  F \right|_{0-}
= -\frac{\tilde{g}_1\tilde{g}_2\Lambda_{QCD}^4}{2\sinh(M\delta)} \sin\theta
\sim 
-\left.\tilde{g}_1\tilde{g}_2 e^{-8\pi^2/g_0^2}\Lambda_{QCD}^4 \sin\theta
\right|_{M\delta >>1}
\ee
This illustrates consistency
of the two pictures, with $\tilde{\kappa} =\tilde{g}_1\tilde{g}_2 $.
Note
the factorization of the overall effect into
the contribution from the two emission vertices 
of the two branes. The current discontinuity is driven
by the overall $\theta$ angle, as the QCD instanton picture 
predicts. 

We now examine the low energy effective Lagrangian including
contributions from both branes and the bulk. 
We substitute our solution
for $F$ back into the action and integrate over $x^5$.
Since our solution is infinitely periodic, 
the brane contributions to the action must
be viewed as  shared between the physical domain $0 \leq x^5 \leq \delta$
and the nearest neighbor image domains. Hence,
each brane contribution to the action in
the physical domain $0 \leq x^5 \leq \delta$ 
receives a 
factor of $1/2$, while the bulk contribution comes from integrating 
eq.(\ref{Lbulk}) over 
$0 \leq x^5 \leq \delta$.  
These contributions
have the same structure and the overall result is:
\begin{eqnarray}
{\cal L}_T & = & \frac{1}{4} (F_{\mu\nu}^a)^2 +
\frac{\Lambda_{QCD}^4 }{4\sinh(M\delta)}\left[
(\tilde{g}_1^2 + \tilde{g}_2^2)\cosh(M\delta)
 +  2\tilde{g}_1 \tilde{g}_2 \cos(\theta ) \right]
\end{eqnarray}
This produces a nontrivial potential in $\theta$, the
residual net CP-angle.  In the large $M\delta$ limit
we obtain:
\be
L \approx \frac{1}{4} (F_{\mu\nu}^a)^2 
+ {\Lambda_{QCD}^4 }\left[\frac{1}{4}(\tilde{g}_1^2 + \tilde{g}_2)
 + \tilde{g}_1 \tilde{g}_2 e^{-M\delta}\cos\theta + O(e^{-2M\delta})\right]
\ee
Note that we have interaction self-energy terms
$\propto (\tilde{g}_1^2 + \tilde{g}_2) $ which have
no $\theta$ dependence. Moreover,
we recover
the $\theta$--potential in QCD $
\propto \Lambda_{QCD}^4
\tilde{g}_1 \tilde{g}_2 \exp(-M\delta) \cos\theta $,
where  the factor
$\exp(-M\delta) = \exp(-8\pi^2/g_0^2)$ now reflects
the duality between the 't Hooft tunneling
amplitude and the t-channel exchange amplitude.
This result implies 
eq.(\ref{338}) identically upon differentiating
wrt $\theta$.
 
Including fermions, the 
emission vertex from brane I must include the `t Hooft
determinant as:
\be
\delta(x^5)\tilde{g}_1 (\Lambda)^p \sqrt{M} \det{\bar{\psi}_L \psi_R}
e^{i\theta_1} F + h.c. 
\ee
If we include $3$ flavors of light quarks, then
$p = -7$.  Normal QCD chiral dynamics
will condense fermions, making a chiral
Lagranagian of pseudoscalar
mesons. We are interested in the
fate of the $\eta'$. hence,
we can ``bosonize'' the `t Hooft determinant
in the usual way, replacing it with the
$\eta'$ field:
\be
\det{\bar{\psi}_L \psi_R} \rightarrow  c\Lambda^9 \exp\left[
\frac{3i\eta'}{\sqrt{6} f_\pi} \right]
\ee
The $\eta'$ mass comes only from
the brane contribution.
If we redefine the monopole $F$ field as $F\rightarrow
\exp({i\theta_1+{3i\eta'}/{\sqrt{6} f_\pi}}) F$ (this
has insignificant effect upon the $\eta'$ kinetic
term), then we
see that the preceding analysis goes through,
yielding a potential which involves the $\eta'$:
\begin{eqnarray}
L & \approx & \frac{1}{4} (F_{\mu\nu}^a)^2 
+ c\tilde{g}_1\tilde{g}_2{\Lambda_{QCD}^4 }\exp(-8\pi^2/g_0^2)\cos( 
 \sqrt{3/2}{\eta'}/f_\pi + \theta  ) 
\end{eqnarray}
This provides the usual resolution to the $U(1)$ problem,
elevating the mass of the $\eta'$.  It also shows that the 
usual $\theta$
problem remains, since other terms in the Lagrangian will
contain $\eta'$ without the $\theta$ angle, and we cannot simply
remove the CP-violation by a shift in the $\eta'$ VEV, e.g.,
the interference terms will lead to a 
nonzero neutron electric dipole moment.

The solution we have 
obtained is the technically appropriate one for the 
specific two
brane model under consideration.  The physics can be
better appreciated in various limits.  
What happens, for example, if we
attempt to remove brane II from the discussion?  Of course,
brane II is serving the role of decoupling the high
energy KK modes, and providing a nonzero monopole mass.
That is, we cannot simply take $\delta \rightarrow \infty$
or we would have a massless monopole, and a 
$4+1$ bulk theory. 
We can consider, however, the
limit $\tilde{g}_2\rightarrow 0$.  Then the $\theta$ angle 
disappears, and also the solution to the $U(1)$ problem.
What has happened?
In this case the physics is controlled by brane I.
The effective Lagrangian is then $\propto \Lambda^2\sqrt{M}
e^{i\theta_1}F + h.c. $.
On brane I, $F$ acquires the VEV, 
$\sim (\Lambda^2/\sqrt{M})e^{i\omega/f_\omega}$, 
where $\omega$ is a free phase factor (a Nambu-Goldstone
boson) that
depends upon $x^\mu$ and $f_\omega \sim \Lambda^2/M$.
Thus the effective
potential in $\omega$ 
is $\sim \Lambda^4\cos(\theta_1 + \omega/f_\omega)$.
The $\theta_1$  angle is now dynamically relaxed to
zero by the $\omega$ field.  This simply restates the
obvious fact that the $\theta_1$ angle is removable
by a phase redefinition of $F$.
In this case $\theta_1$ represents the 
applied or ``thermal cathode current'' of
topological charge into the bulk. The monopole
establishes a compensating field configuration such that
the total current flow, $\propto \sin(\theta_1 + \omega/f_\omega)$
relaxes to zero.  This is analogous to a ``space-charge'' limited
cathode current; the current is quenched by the accumulation
of monopoles in the bulk.  
If we include the $\eta'$ in
this case, we obtain the potential
$\sim \Lambda^4\cos(3\eta'/\sqrt{6}f_\pi + \omega/f_\omega)$.
This is analogous to what happens with an axion (there
are, of course, additional terms which give the $\eta'$ a
mass of order $m_\pi$ without the $\omega$ field in
linear combination). However,
in the case of the axion we have $f_{axion} >> f_\pi$; hence
a see-saw mechanism is always at work in 
axion physics which leaves the $\eta'$ heavy
and makes the axion light.
In the present case,  $f_\omega \sim \Lambda^2/M << f_\pi$, and the
field $\omega$
undergoes large amplitude excursions in a single quantum state,
i.e. $\omega$ is a random phase.  Hence, integrating out
$\omega$ washes out the contribution to the $\eta'$ mass,
and the unwanted fourth $U(1)_A$ Nambu-Goldstone boson returns.

In our two brane model there is a net CP-violation parameterized
by $\theta =\theta_1-\theta_2$, which is physically
analogous to two different thermal cathode currents
on the branes.  The $\omega$ field cannot
relax this linear combination to zero, and an equlibrium monopole
condensate is formed which conducts the net current
$\sim \tilde{g}_1\tilde{g}_2e^{-M\delta}\sin(\theta + 3\eta'/\sqrt{6}f_\pi )$ 
through the bulk.  
The $\eta'$ acquires a mass as this current is
not quenched by $\omega$.

\section{Discussion and Conclusions} 

We have initiated a study of certain gauge field
solitons that occur in higher
dimensional imbeddings of the Standard Model with
gauge fields in the bulk.  We have used the method
of images and the multi-instanton solution in a line
to obtain self-dual solutions consistent with
brane boundary conditions. The monopoles have mass
$8\pi^2M_c/g_0^2$ where $M_c=1/\delta$ is the
inverse compactification scale.

We have considered one application, the reinterpretation
of instantons as the emission/absorption vertices of monopoles 
from the brane into the bulk.  The $U(1)_A$ problem
is then resolved by considering the monopole
condensate which forms in the bulk between branes.
The $\theta$ problem remains, requiring the usual resolution,
e.g.,
in terms of an axion. 
We remark that,
if strong dynamics at high energies lead to a 
large $\sim M_{string}$ condensate 
in $F$ for small instantons $\rho \sim 1/M_c$, then
the phase field $\omega$ may act like an axion
in certain cases. This would be a purely hadronic axion,
i.e., one which couples only to $F^a_{\mu\nu}\tilde{F}^{a\mu\nu}_{QCD}$
but not to
$F_{\mu\nu}\tilde{F}^{a\mu\nu}_{QED}$. This 
may ultimately be natural in these schemes, and  
would alleviate many of the restrictive phenomenological
constraints on $f_{axion}$.

Our reinterpretation of instantons as the exchange of
monopoles at low energies suggests an analogy to the
AdS-CFT correspondence. The, the boundary theory
of $d=0$ pointlike instantons is seen
to be equivalent to the compactified bulk theory
of $d=1$ worldline monopoles at low energies. 

Phenomenologically,
at sufficiently high energies, monopoles 
would be produced
in collisions on the brane.  For example, $g+g \rightarrow F$ is
possible if the $gg$ configuration can produce a
nonzero $F_{\mu\nu}\tilde{F}^{\mu\nu}$. 
Nontopological pair production  $g+g \rightarrow F^\dagger+F$
can also occur.  We expect that the rates may be enhanced at
sufficiently high energy because the modulus $\rho$ must
be summed, and a very large number of states are available (much
like the KK mode production at very high energies of gravitons
where the many KK modes compensate the suppressed coupling).

In the case that $SU(2)$ is identified with $SU(2)_L$,
the monopole charge corresponds to $B+L$, by comparison
with the fermionic current anomalies.  In this case a similar
picture of instanton mediated baryon number violation would
apparently arise, involving corresponding $B+L$ carrying
monopoles in the bulk, and a monopole condensate.  Interestingly,
the $B+L$ violation rates are proportional to $\exp(-M\delta)$,
and could be enhanced if the branes are brought
closer together.  Could this have implications for a theory
of cosmological baryogenesis?

There are  a large number of issues to address
beyond what has been considered here.  Spontaneous 
symmetry
breaking can a lead to a new class of monopoles
involving the (effective) Higgs field.  In the BPS
limit, the construction employed here might
extend to writing down self-dual monopoles
consistent with the brane boundary conditions
by the method of images.
Moreover, the topologies of other objects generalize
similarly, e.g., the 
`t Hooft-Polyakov monopoles become domain walls in one extra
dimension; vortices become 3-branes; etc. 
New objects of different homotopy classes
are expected as well [10]. 

Moreover,
we might ask
what additional dynamics may be involved in
extra-dimensional theories?
For example, it is presumably
straightforward to extend this effective
theory to the Randall-Sundrum model [2], and
we might
expect an intriguing mixing of scales in that case.
Clearly, whatever brane cosmology
emerges, the fate of novel monopoles will be an
issue, requiring some solution such as inflation.
We expect this to be a rich arena for follow-on studies. 
 
\newpage
\noindent
{\bf \Large \bf Acknowledgements}

We wish to thank W. Bardeen, E. Weinberg,
and M. Moshe
for useful comments.
One of us (PR) wishes 
to acknowledge the kind hospitality of the
Fermilab
theory group where this work originated,
and to thank the Aspen Center for Physics
for its kind hospitality during the completion of this work.
\frenchspacing
\vspace*{1cm}

\noindent
{\Large \bf Bibiliography}
\vspace*{0.5cm}
\begin{enumerate}

\item
I. Antoniadis, Phys.Lett. B246 (1990) 377;\\ 
 I. Antoniadis, C. Munoz, M. Quiros 
{\em Nucl. Phys.} {\bf B397} (1993) 515;\\ 
 I. Antoniadis, K. Benakli, M. Quiros
{\em Phys. Lett.} {\bf B331} (1994) 313;\\ 
J.~Lykken, {\em Phys.~Rev.~} {\bf D54}, 3693 (1996);\\
I.~Antoniadis, S.~Dimopoulos and G.~Dvali,
{\em Nucl.\ Phys.\ } {\bf B516}, 70 (1998);\\
N.~Arkani-Hamed, S.~Dimopoulos and G.~Dvali,
{\em Phys.\ Lett.\ } {\bf B429}, 263 (1998);\\
J.~Lykken, L.~Randall,
``The Shape of gravity,''
hep-th/9908076.

\item L.~Randall and R.~Sundrum,
{\em Phys.\ Rev.\ Lett. }  {\bf 83}, 3370 (1999);

\item
K.R.~Dienes, E.~Dudas and T.~Gherghetta,
{\em Phys.\ Lett.} {\bf B436}, 55 (1998);\\ 
K.R.~Dienes, E.~Dudas and T.~Gherghetta,
{\em Nucl.\ Phys.} {\bf B537}, 47 (1999); 
see also: hep-ph/9807522;

\item 
M.~Dine and N.~Seiberg,
{\em Phys.\ Lett.  }{\bf 162B}, 299 (1985);\\
H.-C. Cheng, B. A. Dobrescu and C. T. Hill,
"Electroweak symmetry breaking by extra dimensions",
hep-ph/0004072 and  hep-ph/9912343;\\
H.-C. Cheng, B. A. Dobrescu and C. T. Hill, 
{\em Nucl.Phys.} {\bf B573} 597, (2000);\\
N. Arkani-Hamed, H.-C. Cheng, B. A. Dobrescu and L. J. Hall",
"Self-breaking of the standard model gauge symmetry",
hep-ph/0006238";\\
N.~Arkani-Hamed and M.~Schmaltz,
``Hierarchies without symmetries from extra dimensions,''
hep-ph/9903417.

\item A.A. Belavin, A.M. Polyakov, 
A.S. Shvarts, Yu.S. Tyupkin, 
{\em Phys. Lett.} {\bf B59} 85, (1975) 

\item
G.~'t Hooft,
{\em Phys.\ Rev.\ Lett. }  {\bf 37}, 8 (1976).

\item E. Witten, {\em Phys.\ Rev.\ Lett.} {\bf 38}, 121 (1977).

\item
D. J. Gross, R. D. Pisarski, L. G. Yaffe,
{\em Rev. Mod. Phys} {\bf 53}, 43 (1981); 
Many useful results and references are collected herein.
Our construction bears a strong resemblance to the finite 
temperature considerations discussed in this paper.

\item
R.~Jackiw, C.~Nohl and C.~Rebbi,
{\em Phys.\ Rev.}  {\bf D15}, 1642 (1977).

\item G. Dvali, I. Kogan, M. Shifman, ``Topological Effects
in our Brane World from Extra Dimensions,''
 NYU-TH-00-06-01, hep-th/0006213.

\end{enumerate}
\end{document}